\documentclass[twocolumn,aps,prb]{revtex4-2}

\usepackage{color}
\usepackage{graphicx}

\begin{document}

\title{Persistent current distributions along a $p$-$n$ junction in graphene in a magnetic field}
\author{K. Shizuya}
\affiliation{Yukawa Institute for Theoretical Physics,
Kyoto University,~Kyoto 606-8502,~Japan}

\begin{abstract}
A $p$-$n$ junction, induced in graphene by gating, works to contrast the edge states 
of electrons and holes on each side of it. 
In a magnetic field those edge states carry two species of persistent current, 
which are intimately tied to the edge-mode spectra.
We study how those persistent currents change along each side of the junction as the Fermi level is varied, 
with special emphasis on the electron-hole conjugation property of the Dirac electrons.
A close look is made into the electromagnetic response of the valence band 
filled with negative-energy electrons, or the Dirac sea, 
which as a whole turns out to be electrically inactive 
while showing intrinsic orbital diamagnetism.
Recently, in experiment, it became possible to observe local currents in planar samples
by use of a nanoscale magnetometer.
The $p$-$n$ junctions in graphene and related atomic layers, 
via detection of associated microscopic currents, 
will be a useful platform for exploring many-body quantum phenomena.

\end{abstract} 


\maketitle

\section{introduction}

A long-standing subject related the basis of the quantum Hall (QH) effect~\cite{QHE} 
is the presence of edge states and the current they carry 
in planar systems of  electrons such as semiconductor heterostructures and graphene.
In a magnetic field the edge states support a persistent current circulating along the sample edges, 
as anticipated long ago in connection with the Landau diamagnetism~\cite{Peierls}.  
Theoretically the current flow in QH systems was actively discussed from various angles 
earlier~\cite{Prange,Laughlin,AA,Halperin,MRB, MS,HT,Buttiker,CSG,ThoulessW, AHK,ks_qhe,GV,SSF}. 
In particular,  Geller and Vignale~\cite{GV} pointed out in a model calculation 
that the edge states support two species of current, 
one driven by a local field and another by a density gradient,
that alternate in direction along the sample edges.  

Experimentally, while global transport measurements accumulated 
evidence~\cite{FKHB,WvK,LKK,SBHM,MBA} 
for edge states, 
the microscopic current they carry defied detection for years. Only recently it became possible, 
by use of a SQUID-on-tip nanoscale magnetometer
in experiment of Uri {\it et al.}~\cite{UKBL}, to directly image an alternating pattern 
of edge currents in graphene,
consistent with theoretical expectation.

Graphene, supporting massless Dirac electrons  as charge carriers, 
is an intrinsically many-body system of electrons and holes over the filled valence band 
(acting as the Dirac sea).   
Graphene displays quantum phenomena reflecting the $\lq\lq$relativistic" character of carriers, 
such as Klein tunneling~\cite{KNG} and, in a magnetic field, 
the emergence of zero-energy Landau levels~\cite{NS, Redlich,Jackiw}.

In experiment an edge-like configuration is readily induced
in a sample by electrostatic gating~\cite{HSS,WDM,AL,OJE} 
or about a local charge inhomogeneity~\cite{UKBL}. 
An edge-like configuration, when induced in graphene by gating, provides
an interface that contrasts the edge states of electrons and holes 
in adjacent domains. 
In experiment some notable features of the current flow 
along an electron-hole ($e$-$h$) interface 
(to be called a $p$-$n$ junction below) 
were observed~\cite{UKBL}.

We have earlier examined distinctive features of the two species of edge current in QH systems 
and derived their real-space distributions~\cite{ks_edgeC_W},
which turned out to be intimately tied to the edge-state spectra. 
In the present paper we examine how the current distributions change 
with filling $\sim \epsilon_{\rm F}$ of carriers
along a gate-induced $p$-$n$ junction in graphene, 
with Coulombic many-body effects also taken into account.
Special attention is paid to the $e$-$h$ conjugation property of Dirac electrons, 
that governs the spectra and electromagnetic response of  electrons (of positive and negative energy) 
in two valleys $(K,K')$. 
In particular, it is seen from a linear magnetic response (that detects an orbital magnetic moment) 
that in a magnetic field negative-energy electrons undergo orbital motion 
in the direction opposite to that of positive-energy electrons.
A close look is also taken into the response of the filled  Dirac sea (the valence band), 
which, owing to charge conservation, as a whole 
turns out to be essentially inert electrically while it acts as a diamagnetic medium magnetically. 
This feature naturally leads to a simple picture of persistent currents carried 
by active electrons and holes over the vacuum state on each side of a $p$-$n$ junction.

The paper is organized as follows. In Sec.~II we review and 
refine the derivation of current distributions in graphene 
with the $e$-$h$ conjugation property of the Dirac electron system taken into account. 
In Sec.~III  we examine and simulate persistent-current distributions along a $p$-$n$ junction 
in graphene and, in Sec.~III, discuss how the Coulomb interaction affects them.
Sec.~IV is devoted to summary and discussion.

\section{Graphene}

\subsection{Model and Hamiltonian}

The electrons in graphene are described by two-component spinors 
on two inequivalent lattice sites. 
They acquire a linear spectrum (with velocity $ v_{\rm F} \sim 10^{6}$m/s) 
near the two inequivalent Fermi points $(K,K')$ in momentum space, 
with an effective Hamiltonian of the form~\cite{Semenoff},  
\begin{eqnarray} 
H &=&
\int d^{2}{\bf x}\, \{ \Psi^{\dag}_{+} {\cal H}_{+}\Psi_{+} + \Psi_{-}^{\dag} {\cal H}_{-}\Psi_{-} \},  
\nonumber\\
{\cal H}_{\pm} &=& 
v_{\rm F}\, (\Pi_{x}\sigma^{1}+ \Pi_{y}\sigma^{2})  \pm \delta m\, \sigma^{3}  - eA_{0},
\label{H_GR}
\end{eqnarray}
where $\Pi_{i}= p_{i}+eA_{i}$ and $\sigma^{i}$ denote Pauli matrices.
The Hamiltonians ${\cal H}_{\pm}$ describe electrons 
in two different valleys $(K,K')$ per spin, and 
$\delta m$ stands for a possible sublattice asymmetry; 
we take $\delta m > 0$, without loss of generality.
Valley asymmetry of a few percent is observed 
for high-mobility graphene/hexagonal boron nitride (hBN) devices~\cite{HuntYY,WBEM,PRK}.
For clarity the spin degree of freedom is suppressed, with its splitting set to zero.

Let us place graphene in a uniform magnetic field $B_{z}=B>0$ 
with potentials $(A_{x}, A_{y}) = (-By, 0)$
and, to study the electromagnetic response of the system, 
introduce also weak external potentials ${\bf a} = (a_{x}, a_{y})$.
We write ${\bf a}$ in a dimensionless form 
${\bf v} = e\ell\,  {\bf a}$, using magnetic length  $\ell = 1/\sqrt{eB}$,
and set $v({\bf x}) = e \ell\, \{a_{y}({\bf x}) + ia_{x}({\bf x}) \}/\sqrt{2}$.
The Hamiltonian ${\cal H}_{+}$ in valley $K$ is written as
\begin{equation}
{\cal H}_{+} = \omega_{c} \left(
\begin{array}{cc}
\mu & -Z \!-i v({\bf x}) \\
-Z^{\dag} \!+iv^{\dag}({\bf x})& -\mu \\
\end{array}\!
\right) - e A_{0}({\bf x}), 
\label{Hv_gr}
\end{equation}
where $\omega_{c} \equiv \sqrt{2}\, v_{\rm F}/\ell$ and $\mu \equiv \delta m/\omega_{c}$;
$Z= (Y+ iP)/\sqrt{2}$ and $Z^{\dag}= (Y- iP)/\sqrt{2}$ in terms of 
$Y= (y-y_{0})/\ell$ and $P =- i\ell \partial_{y}$ with $[Y, P]=i$ 
and momentum $p_{x}$ or $y_{0} \equiv \ell^2 p_{x}$.
External potentials $\{v({\bf x}), A_{0}({\bf x})\}$ are taken to vary slowly in space 
(and time $t$, left implicit).
For a sample homogeneous in one direction $\{x\}$, 
a simple and practical choice is to take them to vary only in the width direction $\{y\}$.
One can then use a potential $a_{x}(y)$ to detect an $x$-averaged current
$j_{x}(y) = (1/L_{x}) \int dx\,  j_{x}(x,y)$ (with $L_{x} = \int dx$) 
driven by a local field 
$E_{y}(y) = -\partial_{y}A_{0}(y) - \partial_{t}a_{y}(y)$,
while probing a magnetic response of electrons 
via a field 
$b_{z}^{\rm ({\bf a})} \equiv  \nabla \times {\bf a} = \partial_{x}a_{y} -\partial_{y}a_{x}
 \rightarrow  -\partial_{y}a_{x}(y)$
 normal to the sample. We adopt such a choice later.

For $v=A_{0}=0$, the electron spectrum forms an infinite tower 
of Landau levels of energy 
\begin{equation} 
\epsilon_{n} =\omega_{c} e_{n} \ {\rm and}\ \ e_{n} \equiv s_{n} \sqrt{|n|+\mu^{2}}
\end{equation}
in each valley (with $s_{n}\equiv  {\rm sgn}[n] = \pm1$), 
labeled by integers $n = (0,\pm 1, \pm2, \dots)$ and
$y_{0}=\ell^2 p_{x}$, of which only the $n=0$ (zero-mode) levels split in the valley
(hence to be denoted as $n=0_{\mp}$), 
\begin{equation}
\epsilon_{0_{\mp}}= \mp \delta m = \mp \omega_{c}\,  \mu   \ \ {\rm for}\ K/K'.
\end{equation}
Thus, for each integer $|n| \equiv N = 0,1,2, \cdots$ 
(we use capital letters for the absolute values),
there are in general two modes with $n=\pm N$ (of positive/negative energy) 
per valley and spin, apart from the $n=0_{\pm}$ modes. 

The eigenmodes $(n, y_{0})$ in valley $K$ are written as~\cite{KS_LWGS}
\begin{equation}
\langle {\bf x}|\Phi^{n}_{y_{0}}\rangle 
= \big( 
\langle {\bf x}  | N\! -\!1, y_{0} \rangle\, b^{n}, \langle {\bf x} | N, y_{0} \rangle\, c^{n} 
\big)^{\rm t},
\label{psi_n}
\end{equation}
with $N=|n|$.  
Here $\langle {\bf x}|N, y_{0} \rangle = \langle x|y_{0}\rangle\,  \phi_{N}(y-y_{0})$
consist of plane waves 
$\langle x|y_{0}\rangle = e^{ix\,y_{0}/\ell^2}/\sqrt{2\pi \ell^2}$
of momentum $p_{x}$
and the harmonic-oscillator wave functions  
\begin{equation}
\phi_{N}(y)
= e^{-{1\over{2}} (y/\ell)^2}\, H_{N}(y/\ell)/\sqrt{N!\, 2^{N}\sqrt{\pi}\,  \ell}.
\end{equation}
They are  the wave functions of conventional electrons (of quadratic dispersion), 
labeled by orbitals
$N=0,1,2, \cdots$ and center position $y_{0}= \ell^2 p_{x}$. 
The factors $(b^{n}, c^{n})^{\rm t}$ are given 
by the orthonormal eigenvectors of ${\cal H}_{+}$ in Eq.~(\ref{Hv_gr}) 
with  $(v, A_{0}) \rightarrow 0$ and $(Z, Z^{\dag}) \rightarrow \sqrt{N}$.
In explicit form,
\begin{equation}
(b^{n}, c^{n}) 
= {1\over{\sqrt{2}}}\, \left( \sqrt{1+ {\mu\over{e_{n}}} }, -s_{n} \sqrt{1- {\mu\over{e_{n}}} }\right)
\label{bc_muzero}
\end{equation}
for $n= \pm N$, and $(b^{0_{-}}, c^{0_{-}})=(0, 1)$.

In the $|N,y_{0}\rangle$ basis, the coordinate 
$\langle N, y_{0}|{\bf x}|N', y'_{0}\rangle= ({\bf X}^{NN'} + \delta^{NN'} {\bf r} ) \delta (y_{0} -y'_{0})$
is split into the relative coordinate ${\bf X} =(X_{x}, X_{y})  = \ell (P, Y)$
and center coordinate ${\bf r} = (i\ell^2 \partial_{y_{0}}, y_{0})$, 
with  $[X_{x}, X_{y}] =-i\ell^{2}$, $[r_{x}, r_{y}] =i\ell^{2}$ and $[X_{i},r_{j}]=0$.
Here 
${\bf X}$ or $(Z, Z^{\dag})$ become 
familiar numerical matrices in the harmonic-oscillator basis  $\{ |N\rangle \}$;
$Z^{MN} \equiv \langle M|Z|N\rangle = \sqrt{N}\, \delta^{M, N-1}$ 
and $[Z, Z^{\dag}]=1$.

Correspondingly, in this basis, a function of ${\bf x}$, 
e.g., $v({\bf x})$, becomes a matrix in orbital labels,
$\langle M,y_{0}|v({\bf x})|N,y'_{0}\rangle = [v({\bf X+ r})]^{MN} \delta (y_{0}-y'_{0})$, 
and its matrix portion is isolated, by use of a Fourier transform as 
\begin{equation}
[v({\bf x})]^{MN}= \sum_{\bf p} [e^{i{\bf p \cdot (X+r)}}]^{MN} v_{\bf p}
= [e^{i{\bf p \cdot X}}]^{MN}v({\bf r}),
\end{equation} 
where $v({\bf r}) \equiv \sum_{\bf p}v_{\bf p} e^{i{\bf p \cdot r}}$; 
$\sum_{\bf p} \equiv \int d^{2}{\bf p}/(2\pi)^{2}$.
In the last line, we regard ${\bf p}$ as a derivative $-i\nabla$ acting on $v({\bf r})$;
$ip_{y}v({\bf r})$, e.g., stands for $\partial_{y}v({\bf x})$ with ${\bf x} \rightarrow {\bf r}$.
Both $e^{i{\bf p \cdot X}}$ and $e^{i{\bf p \cdot r}}$ obey the $W_{\infty}$ algebra~\cite{GMP}. 
One can rewrite 
\begin{equation}
e^{i{\bf p \cdot X}} = e^{i  \ell p Z^{\dag} + i  \ell p^{\dag} Z}
=   \gamma_{\bf p}\, e^{i\ell pZ^{\dag}} e^{i\ell p^{\dag}Z} \equiv  \gamma_{\bf p} f_{\bf p}
\end{equation}
with  $\gamma_{\bf p} =e^{-{1\over{4}}\ell^2 {\bf p}^2}=e^{-{1\over{2}}\ell^2 p^{\dag}p}$,
where
$p \equiv (p_{y} + ip_{x})/\sqrt{2}$, or $ip = (\partial_{y} + i\partial_{x})/\sqrt{2}\equiv \partial$
and $ip^{\dag} =  \partial^{\dag}$.  
Then  $v({\bf x})$ is naturally expanded in a normal-ordered series of $(Z^{\dag}, Z)$ 
and in multipoles of $v({\bf r})$.
Note that such operations 
as $[Z, v({\bf x})] = \ell \partial v({\bf x})$ and 
$[Z^{\dag}, v({\bf x})] = -\ell \partial^{\dag} v({\bf x})$ now hold.

The matrix elements
$f^{MN}_{\bf p} = \langle M| f_{\bf p}|N \rangle$ 
are expressed in terms of the associated Laguerre polynomials, 
\begin{equation}
f^{MN}_{\bf p} = \sqrt{N!/M!}\,  (i\ell p)^{M-N} L^{M-N}_{N}(\ell^{2}p^{\dag}p).
\label{fmn}
\end{equation}
They are also related to the Fourier integrals of quadratic harmonic-oscillator wave functions,
\begin{equation}
[e^{i{\bf p} \cdot {\bf X}}]^{MN}
= \int_{-\infty}^{\infty} \!\! dy\,  e^{i p_{y} y} 
\phi_{M}(y- \Delta_{p})\,  \phi_{N}(y + \Delta_{p}) ,
\label{Fourier_fMN}
\end{equation} 
where $\Delta_{p} = {1\over{2}}\ell^2 p_{x}$.
In the spinor basis $\{ |\Phi^{n}_{y_{0}}\rangle\}$ 
one can write 
$[e^{i{\bf p} \cdot {\bf X}}]^{mn} = \gamma_{\bf p} \, g^{mn}_{\bf p}$ 
with 
$g^{mn}_{\bf p} \equiv b^{m} f^{M-1,N-1}_{\bf p} b^{n} + c^{m} f^{MN}_{\bf p} c^{n}$.

Let us now turn on $(v, A_{0})$ and 
expand $\Psi_{+}({\bf x}) =\sum_{y_{0},n} \langle {\bf x}| \Phi^{n}_{y_{0}}\rangle\, \psi_{n}(y_{0})$ 
in the basis $\{ |\Phi^{n}_{y_{0}}\rangle \}$.
The one-body Hamiltonian $H_{+}$ of valley $K$ is then written as~\cite{KS_LWGS}
\begin{eqnarray}
H &=&\int\! d y_{0}\,  \psi_{m}^{\dag}(y_{0}) {\cal H}^{mn} \psi_{n}(y_{0}),
\nonumber\\
{\cal H} 
&=& \epsilon + \omega_{c} i (c\,  v^{\dag}\, b  -b\,v\, c)-  e\, (b\, A_{0}\, b +  c\,A_{0}\,c), \ \ \ 
\label{H_gr}
\end{eqnarray}
where $\epsilon$ stands for diagonal spectra $\epsilon_{n}\, \delta^{mn}$.
Here ${\cal H}$ represents a matrix ${\cal H}^{mn}$ 
in orbital labels $(m,n)$ which now run over all integers $(0_{-}, \pm1,\pm2, \cdots)$; 
in what follows we adopt such matrix notation and 
frequently suppress summation over repeated labels.
In  ${\cal H}$ we have introduced condensed notation: 
For ${\cal H}^{mn}$ we interpret, e.g., 
\begin{eqnarray}
b\, v\, c  &\rightarrow&    b^{m}\, [v({\bf x})]^{M-1,N} c^{n},
\nonumber\\
c\, v^{\dag}\, b  &\rightarrow&    c^{m}\, [v^{\dag}({\bf x})]^{M,N-1} b^{n},
\nonumber\\
c\,A_{0}\,  c\,   &\rightarrow&\,  c^{m}\, [A_{0}({\bf x})]^{M,N}\, c^{n}, 
\label{condensed_nt}
\end{eqnarray}
where $[v({\bf x})]^{MN} =  [e^{i{\bf p \cdot X}}]^{MN}v({\bf r})$, etc.

\subsection{Electron-hole conjugation and the Dirac sea}

Our present focus is on current distributions. 
The electric current ${\bf j} =(j_{x}, j_{y}) = - \delta H/\delta {\bf a}$ 
is read from the Hamiltonian $H$ through terms linear in $a_{j}$.
The external probes $(v,A_{0})$, upon introduction into ${\cal H}$,
induce mixing of Landau levels.
Such level mixing is resolved via diagonalization of ${\cal H}^{mn}$ in orbitals 
by a suitable unitary transformation $U \equiv U^{mn}({\bf r})$,
with
$\hat{\psi}_{m}(y_{0}) = U^{mn} \psi_{n} (y_{0})$ and 
$\hat{\cal H} =U \{ {\cal H} - i\partial_{t}\} U^{-1}$.
One can then derive from 
\begin{equation}
\hat{H} 
=\int\! d y_{0}\,  \hat{\psi}_{n}^{\dag}(y_{0}) \, \hat{\cal H}^{nn}({\bf r})\, \hat{\psi}_{n}(y_{0})
\end{equation} 
the level spectra and current ${\bf j}$ diagonal in orbitals $\{n\}$.

For clarity of exposition let us adopt probes $v(y)$ and $A_{0}(y)$ 
that vary only in the width direction $y =y_{0} + \ell Y$ of a  sample. 
Then the center position $y_{0}= \ell^2 p_{x}$ becomes a good quantum number and  
$\hat{\cal H}^{nn}({\bf r}) \equiv {\cal E}_{n}(y_{0})$ denote the spectra of 
the eigenmodes $\hat{\psi}_{n}(y_{0})$.

Let us first look into some general features of  the spectra ${\cal E}_{n} \equiv {\cal E}_{n}(y_{0})$.
[Hereafter we refer, when necessary, to a specific valley by a superscript $(K,K')$.] 
One can pass to another valley $K'$ by simply setting $\mu\rightarrow - \mu$ 
since ${\cal H}_{-} =  {\cal H}_{+}|_{-\mu}$ holds, 
 where  ${\cal O}|_{-\mu}$ signifies reversing the sign of $\mu$ in ${\cal O}$.
 One thus has the relation 
  ${\cal E}^{K'}_{n} = {\cal E}^{K}_{n}|_{-\mu}$.  
 Let us further note the unitary equivalence of ${\cal H}_{\pm}$,
\begin{equation}
\sigma^{3}\, {\cal H}_{-}\sigma^{3} = -{\cal H}_{+}|_{-A_{0}},\ \ 
 \sigma^{3}\, {\cal H}_{+}\sigma^{3} = -{\cal H}_{+}|_{-A_{0}, -\mu},
\end{equation}
which, in view of the discreteness of the level spectra, implies the following relations
\begin{eqnarray}
{\cal E}^{K'}_{n} &=& -{\cal E}^{K}_{-n}|_{-A_{0}} 
= {\cal E}^{K}_{n}|_{-\mu} 
\ \ {\rm for\ } n\not = 0,
\nonumber\\
{\cal E}^{K'}_{0_{+}} &=&   -{\cal E}^{K}_{0_{-}}|_{-A_{0}}
= {\cal E}^{K}_{0_{-}}|_{-\mu}.
\label{EKandEKp}
\end{eqnarray}
Via this electron-hole conjugation,  
all the level spectra in two valleys are fixed 
by $\{ {\cal E}_{n}^{K}; n= 0_{-},1,2, \cdots \}$ alone.

These $e$-$h$ conjugated spectra reveal
how the Dirac electrons of level $n=\pm N$ (i.e., of positive/negative energy) 
respond to external probes $(v, A_{0})$.
(i)~Their response of $O(v) \ni O(v_{x}), O(\partial_{y}v_{x})$ is opposite in sign for $n=\pm N \not=0$.
This implies that the current associated with the cyclotron motion of electrons changes sign 
for $n=\pm N$.  Thus, in a magnetic field $B$, positive- and negative-energy electrons 
orbit in mutually opposite directions. 
Each negative-energy electron therefore induces a microscopic circulating current 
of a {\it paramagnetic} nature 
in contrast to a normal diamagnetic current associated with each positive-energy electron.
The current ${\bf j}$ of level $n$, in general, differs in the sign of $\mu$ between valleys $K$ and $K'$; 
 actually,  there arises no term odd in $\mu$, as we later see from $\kappa^{nn}(\xi)$ in Eq.~(\ref{Vnr}).
 Hence the circulating component of current ${\bf j}$ of level $n$ takes the same form in both valleys.  
(ii)~The electrons of zero-mode levels $n=0_{\mp}$ 
show no $O(v)$ response for $\mu \rightarrow 0$ 
[and also for $\mu\not=0$ with $\kappa^{00}(\xi)=0$  in Eq.~(\ref{Vnr})].
They thus make no cyclotron motion.

(iii)~The $O(v A_{0})$ response, bilinear in $(v, A_{0})$, 
detects a portion of current ${\bf j}$ driven by a local potential $A_{0}({\bf x})$, i.e., 
a Hall current. This response is the same for positive- and negative-energy electrons 
of levels $n^{K}=-n^{K'}$ in different valleys, so is the associated Hall current.
The meaning of Eq.~(\ref{EKandEKp}) becomes clearer 
when one passes to an electron-hole picture in Eq.~(\ref{hole_e_spectra}) later.

In practice, to derive the level spectra and current ${\bf j}$, 
it is sufficient to diagonalize ${\cal H}^{mn}$ 
to $O(v)$,  $O(A_{0})$ and $O(vA_{0})$ 
by constructing the unitary transformation $U^{mn}$ perturbatively.
As for the $O(vA_{0})$ response, 
we now retain up to a single derivative $\partial A_{0} \sim -E_{y}$ for $A_{0}(y)$ 
[while keeping full multipoles of $v(y)$], 
assuming its gentle spatial variations. 
Appendix A presents in some detail the perturbative construction of $\hat{\cal H}$.

Actually diagonalization of ${\cal H}^{mn}$ was made earlier 
by use of a method of gauge theory~\cite{ks_edgeC_W}, 
which considerably simplifies actual calculations. 
One can regard external probes $(v, A_{0})$ as a gauge field that induces $U(\infty)$ 
(or $W_{\infty}$) rotations of infinitely many Landau levels, 
and make use of a special $W_{\infty}$ gauge transformation 
to carry out diagonalization in such a manner 
that electromagnetic gauge invariance is kept manifest.

The spectra ${\cal E}^{K}_{n}({\bf r})$ diagonal to $O(v)$, $O(A_{0})$ 
and $O(v\partial A_{0})$ are written as~\cite{ks_edgeC_W, fn} 
\begin{eqnarray}
{\cal E}^{K}_{n}({\bf r}) &=& \epsilon_{n}[{\bf r}]  + V_{n}[{\bf r}],
\label{EnKone}
\\
\epsilon_{n}[{\bf r}] &=&\epsilon_{n}+ g^{nn}(\xi)\, \gamma_{\bf p}  W({\bf r})\,
\nonumber\\
&\approx& \epsilon_{n}   + W({\bf r}) 
+  O(\nabla^2 W), 
\\
V_{n}[{\bf r}] &=& - \omega_{c}\, \kappa^{nn} (\xi) \,\gamma_{\bf p}\, b_{z}({\bf r}) 
\nonumber\\
&&+ \{g^{nn}(\xi) \gamma_{\bf p}\,  {\bf v}({\bf r})\} \times e \ell {\bf E}({\bf r})
\nonumber\\
&& 
- \{h^{nn}(\xi) \gamma_{\bf p} \nabla b_{z}({\bf r})\} \cdot e \ell {\bf E}({\bf r}),\ \ \ 
\label{Vnr}
\end{eqnarray}
with a magnetic probe 
$b_{z}({\bf r}) \equiv  \ell\,  \nabla \times {\bf v}({\bf r}) 
= e\ell^2  \nabla \times {\bf a}$ and a local potential $W({\bf r})= -eA_{0}({\bf r})$ for electrons; 
${\bf E} = (E_{x}, E_{y}) = -\nabla A_{0}- \partial_{t}{\bf a}$; 
${\bf v}\times {\bf E} = v_{x}E_{y} - v_{y}E_{x}$.
Here
\begin{eqnarray}
 g^{nn}(\xi) &\equiv& b^{n}  b^{n} f^{N-1,N-1}_{\bf p} + c^{n} c^{n} f^{NN}_{\bf p},  
 \nonumber\\
&=& b^{n}  b^{n} L_{N-1}(\xi) + c^{n} c^{n} L_{N}(\xi),
\nonumber\\
\kappa^{nn}(\xi) &\equiv& b^{n}\, c^{n}\,\{f_{\bf p}^{N,N-1}/(i\ell p)\} 
=  - L_{N-1}^{1}(\xi)/(2e_{n}),
\nonumber\\
h^{nn}(\xi) &=& L^{2}_{N-1}(\xi)  -L^{2}_{N-2}(\xi) = L_{N-1}^{1}(\xi), 
\end{eqnarray}
and $\gamma_{\bf p}= e^{-{1\over{4}} \ell^2 {\bf p}^2}= e^{-{1\over{2}} \xi}$ 
are functions of a derivative 
$\xi =-\ell^{2} \partial^{\dag}\partial = - {1\over{2}}\ell^{2} \nabla^2$
acting on $v({\bf r})$; $g^{nn}(0) =1$, $ \kappa^{nn}(0)=-N/(2e_{n})$ 
and $h^{nn}(0) =N$.
In the above, for generality, we have retained  ${\bf r} = (i\ell^2\partial_{y_{0}}, y_{0})$.
For static potentials $v(y)$ and $A_{0}(y)$ of our present choice, 
one can set
${\bf r} \rightarrow y_{0}$, $v({\bf r}) \rightarrow v(y_{0})$, 
$\xi  \rightarrow - {1\over{2}} \ell^{2} (\partial_{y_{0}})^{2}$, etc., in ${\cal E}^{K}_{n}({\bf r})$. 
One can further simplify $v(y_{0}) \rightarrow i v_{x}(y_{0})/\sqrt{2}$,
$b_{z}(y_{0}) \rightarrow - \ell\, \partial_{y_{0}}v_{x}(y_{0}) \equiv -\ell v'_{x}(y_{0})$, etc.,
since $v_{y}(y_{0})$ trivially disappears from ${\cal E}_{n}^{K}(y_{0})$.

From the $b_{z}$ term in $V_{n}[{\bf r}]$
one can read off an orbital magnetic moment of an electron, 
$\hat{m}_{n} = e\ell^2\,  \omega_{c}\, \kappa^{nn}(0)$ 
or
\begin{equation}
\hat{m}_{n} = - e \ell^2 \omega_{c} N/(2 e_{n}) \approx  - s_{n}\,e\ell v_{\rm F}\sqrt{N/2}, 
\label{MagM_n}
\end{equation}
which, as noted above, changes sign for $n=\pm N$ and vanishes for $n=0_{\mp}$. 
In cyclotron motion the negative-energy electrons (of level $n =-1,-2, \cdots$) 
orbit  in the opposite direction to positive-energy ones.

 \begin{figure}[bpt]
\begin{center}
\includegraphics[scale=.8]{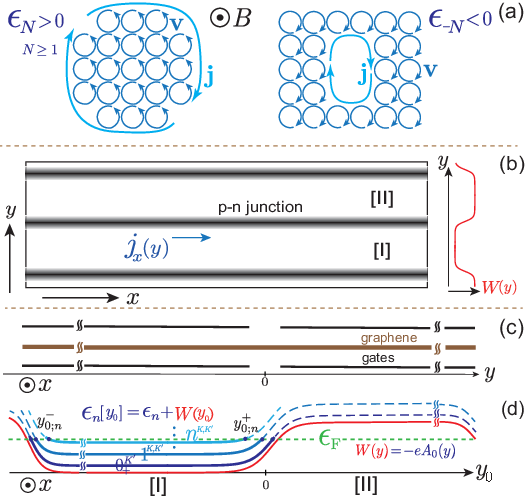}
\end{center}
\vskip-.7cm
\caption{
(a)~Semiclassical picture of a hole excitation (right) in the sea of
negative-energy electrons, which, in cyclotron motion, 
orbit oppositely to normal positive-energy ones (left).
Along an empty domain (a hole excitation) in the sea, 
a macroscopic current survives and circulates in the same direction as that of normal electrons.  
(b)~Schematic of a graphene sample, 
divided into some domains across junctions (shaded areas) by electrostatic gating.  
(c)~Cross-section view of the sample.
(c)~A static potential wall $W(y)  = -e A_{0}(y)$ (red curve) is formed around $y \sim 0$ by gating. 
Electrons of some lower levels are accommodated in domain [I] by the wall 
when the Fermi energy $\epsilon_{\rm F}$ is set to cross the wall. 
}
\end{figure}

It is enlightening to interpret this situation intuitively in a semi-classical picture:
Associated with a normal electron in cyclotron motion is a microscopic circulating current. 
This current cancels out locally in a densely populated domain of electrons 
while it leaves a macroscopic current ${\bf j}$ circulating along the periphery and 
magnetization $\hat{m}_{n}\bar{\rho}$ inside,  as illustrated in Fig.~1(a) (left).
Let us now imagine a graphene sample filled with negative-energy electrons 
orbiting in the opposite direction to normal and suppose that an empty domain is formed in it anew. 
Then, as depicted in Fig.~1(a) (right), 
a macroscopic circulating current ${\bf j}$ survives along the periphery of the empty domain 
and flows in the same direction as that of normal electrons. 
In the hole picture [of Eq.~(\ref{hole_e_spectra}) shown later], 
this current is attributed to the cyclotron motion of holes of positive charge and energy, 
moving oppositely to electrons.

There are infinitely many Landau levels in graphene.
In particular, the valence band, or the Dirac sea, consists of 
all filled negative-energy levels,
i.e., levels $\{n \}$ with $n \le 0_{-}$ in valley $K$ and $n \le -1$ in $K'$, and 
has the energy density (per unit area and spin),
\begin{eqnarray}
\epsilon_{\rm vac}^{B} 
&=& \bar{\rho}\, \Big[   {\cal E}_{0_{-}}^{K}(y_{0})  
+ \sum_{n\ge 1}^{N_{\rm cut}} \{ {\cal E}_{-n}^{K}(y_{0}) + {\cal E}_{-n}^{K'}(y_{0}) \} \Big],
\label{Evac_zeroB}
\end{eqnarray}
with its depth cut off at level $n= -N_{\rm cut}$;  $\bar{\rho} = 1/(2\pi \ell^2)$.

The filled sea in each valley 
has a nontrivial response to external probes $(v, A_{0})$.
One may naively expect an infinitely large vacuum current 
since each filled level in the sea carries 
a Hall current $j_{x}$, driven by a local field $E_{y}$, 
with conductance $\sigma_{xy} = - e^2/h$ per level. 
This, of course, is not the case.
With careful handling of charge conservation at the bottom of the sea 
the bottom level turns out to carry a large amount of vacuum current, 
which leaves only a tiny amount of response, 
characterized by a half unit of vacuum Hall conductance, 
$\sigma^{{\rm vac}}_{xy}|^{K/K'} = \mp (e^2/4\pi \hbar) = \mp {1\over{2}} e^2/h$
per spin in each valley.
See Appendix A for details.   
Such a vacuum response, also known as a parity anomaly~\cite{Redlich},  
has topological origin in the nonzero index~\cite{NS,Jackiw} 
of the Dirac Hamiltonian $H_{\pm}$.
The filled Dirac sea (with $K+K'$) as a whole thus has no response
to weak perturbations $(v,A_{0})$, 
and stays as a $\lq\lq$vacuum" state $|0\rangle$ of filling factor $\nu=0$.

The vacuum state still has a nontrivial response to a uniform field $B$
through
\begin{equation}
\epsilon_{\rm vac}^{B}|_{v=A_{0}=0} \approx -\bar{\rho}\, 
\Big\{ 2\sum_{n=1}^{N_{\rm cut}}  \epsilon_{n} + |\delta m| \Big\},
\label{E_B_vac}
\end{equation}
which is to be compared with the $B\rightarrow 0$
vacuum energy 
$\epsilon^{B=0}_{\rm vac} = - 2 \sum_{\bf k} \sqrt{ v_{\rm F}^2\, {\bf k}^2 + (\delta m)^2}$,
with 
the Fermi momentum $k_{\rm F}$ chosen to support the same number of negative-energy states, 
$N_{\rm sea} = k_{\rm F}^2/(4\pi) = (N_{\rm cut}+{1\over{2}})\,  \bar{\rho}$ per area and per valley.
The deviation $\epsilon_{\rm vac} =\epsilon^{B}_{\rm vac} -\epsilon^{B=0}_{\rm vac}$,
finite for $N_{\rm cut}\rightarrow \infty$, 
is a real physical energy shift that the vacuum state experiences 
when $B$ is turned on gradually~\cite{ks_edgeC_W},
\begin{equation} 
\epsilon_{\rm vac} = \bar{\rho}\, \omega_{c} \left\{ - 2\, \zeta(-1/2) - |\mu|  + O(\mu^2)\right\} >0,
\label{E_vac}
\end{equation}
with a zeta function $-\zeta(-1/2) = \zeta(3/2)/(4\pi) \approx 0.2079$;
see Appendix B for details.
This expression for $\epsilon_{\rm vac} \propto B^{3/2}$ 
was also derived earlier in thermodynamic calculations~\cite{McC,SGB,GGC}.
Suppose now that this $\epsilon_{\rm vac}$ effectively comes from "carriers" 
distributed with density $\bar{\rho}$ per valley.  
Then such carriers will have an orbital magnetic moment 
$- (\partial/\partial B)(\epsilon_{\rm vac}/2\bar{\rho}) 
={1\over{2}}e\ell^{2}\omega_{c}\{\zeta (-{1\over{2}}) + \cdots\} < 0$  per carrier. 
As noted in Eq~(\ref{MagM_n}), 
negative-energy electrons acquire a paramagnetic orbital moment 
$\hat{m}_{n=-N} \approx {1\over{2}} e \ell^2 \omega_{c} \sqrt{N}  >0$  
per carrier. 
Accordingly, 
in the physical energy shift $\epsilon_{\rm vac}^{B} - \epsilon^{B=0}_{\rm vac}$, 
infinite orbital magnetization, naively expected for a filled sea, 
\begin{equation}
\sum_{N=0}^{\infty} \sqrt{N}  \rightarrow 
\zeta (-{\textstyle{1\over{2}} } ) = -\zeta({\textstyle{3\over{2}}}) /(4\pi) \approx -0.2079  <0,
\end{equation}
turns into a small value of opposite sign.
The vacuum state in graphene, with its energy increasing with $B$, 
thus has intrinsic {\it diamagnetic} magnetization 
\begin{equation}
M_{\rm vac} = - \partial \epsilon_{\rm vac}/\partial B 
\approx  3(e\omega_{c}/2\pi)\,  \zeta(-{\textstyle{1\over{2}}}) <0
 \end{equation}
per spin, with no active charge carriers.
This vacuum magnetization is a manifestation of orbital magnetic moments 
of negative-energy electrons regularized over the filled Dirac sea.
In graphene orbital magnetization increases rapidly 
when one approaches the $\nu=0$ vacuum state by increasing $B$,
as observed recently in experiment~\cite{BWFP}.

Clearly it is natural to measure everything relative to this neutral state 
and specify the negative-energy states as holes, 
with fields 
$\hat{\chi}_{n}(y_{0}) \equiv \hat{\psi}_{-n}^{\dag}(y_{0})$ ($n=0_{+}, 1, 2, \cdots$,
in valley $K$). 
Let us isolate the $\epsilon_{\rm vac}$ term and write the total Hamiltonian as 
$\hat{H} = \hat{H}^{K}+ \hat{H}^{K'} +  L_{x}\int dy_{0}\, \epsilon_{\rm vac}$,
where $\hat{H}^{K}$, e.g., now stands for 
\begin{equation}
\hat{H}^{K} = \int\! dy_{0} 
\Big\{
\sum_{n =1}^{\infty} \hat{\psi}_{n}^{\dag}  {\cal E}_{n}^{K}(y_{0}) \hat{\psi}_{n} 
+ \sum_{n = 0_{+}}^{\infty} \hat{\chi}_{n}^{\dag}  {\cal E}_{n}^{{\rm hole};K}(y_{0}) \hat{\chi}_{n} 
\Big\},
\end{equation}
with 
${\cal E}^{{\rm hole};K}_{n} \stackrel{n > 0}{=} -{\cal E}^{K}_{-n}(y_{0})$.
In view of Eq.~(\ref{EKandEKp}), 
holes in one valley behave like electrons in another valley 
with an inverted potential $A_{0}$,
\begin{eqnarray}
&& \{  {\cal E}^{{\rm hole};K}_{n} , {\cal E}^{{\rm hole};K'}_{n} \}
= \{ {\cal E}^{K'}_{n}, {\cal E}^{K}_{n}\}|_{-A_{0}} \ \ (n\ge 1),
\nonumber\\
&&{\cal E}_{0_{+}}^{{\rm hole};K} = {\cal E}^{K'}_{0_{+}}|_{-A_{0}}.
\label{hole_e_spectra}
\end{eqnarray}
Clearly, holes have positive charge $e>0$ and move oppositely to electrons. 

\subsection{Two species of edge current}

To study the current in graphene 
one may now simply pass to a picture of electrons and holes near the vacuum state 
without handling the presence of the Dirac sea. 
Figure 1(b) depicts a Hall-bar sample we consider;
Fig.~1(c) shows its cross-section view.   
It extends homogeneously in the $x$ direction 
while, in the width direction $\{y\}$, it is divided into two or more domains 
by gating, as done in experiment~\cite{UKBL}.  
In each domain the static potential $W(y)= -eA_{0}(y)$ (drawn with a red curve) 
is taken to be flat except for the edge regions 
where $W(y)$ connects adjacent domains smoothly.
The {\it wall} $W(y)$ locally raises the level spectra $\epsilon_{n}[y_{0}]$ 
and the neutral surface of the filled Dirac sea,
and easily acts to form a junction of active electrons and holes.
Let us, for the moment, focus on domain [I]  in Fig.~1(d)
and examine how it supports an edge current $j_{x}(y)$ along the wall. 
We suppose that $W(y)$ stays 0 in the interior 
and rises in the lower and upper edge regions so that it 
accommodates active electrons of some lower levels
$\{n\}=\{0_{+}^{K'}, 1^{K,K'}, 2^{K,K'}, \cdots\}$ in the domain.

Varying 
$\hat{H}^{K} =\int\! dy_{0} \sum_{n} \hat{\psi}_{n}^{\dag}  {\cal E}_{n}^{K}(y_{0}) \hat{\psi}_{n}$ 
with respect to $v_{x}(y_{0})$ yields 
a current $j_{x}[y_{0}] =- e\ell\,  \delta \hat{H}^{K}/\delta v_{x}(y_{0})$ 
in the eigenmode space $(n, y_{0})$.
To convert it to a real-space current $j_{x}(y)$, 
which is observable in experiment, one first has to 
switch from $v_{x}(y_{0}) [= \sum_{\bf p}(v_{x})_{\bf p}\, e^{i{\bf p \cdot r}}]$ 
to a real-space potential $v_{x}(y)$, 
i.e., set $v_{x}(y_{0}) =  \int\! dy\, v_{x}(y)  \sum_{q} e^{iq (y_{0}-y)}$ 
in ${\cal E}_{n}^{K}(y_{0})$;  $\sum_{q} = \int dq/(2\pi)$.
In view of Eq.~(\ref{Fourier_fMN}), one thereby encounters some quadratic forms 
of harmonic-oscillator functions $\{ \phi_{N}(y) \}$, 
\begin{equation}
\sum_{q}e^{-i q y -{1\over{4}} q^2} L_{N}({\textstyle{1\over{2}}} q^2) = |\phi_{N}(y)|^2, {\rm etc}.
\label{Fourier_Ln}
\end{equation}
The real-space current is thereby written as
\begin{eqnarray} 
j_{x}(y) &=& -{e\ell^{2}\over{L_{x}}}\sum_{n}\int dy_{0}\, 
I_{n} (y-y_{0})\, \hat{\rho}_{n}(y_{0}),
\nonumber\\
I_{n}(y) &=& -\omega_{c}\partial_{y}{\cal M}_{n}(y)  + W'(y_{0})\, \hat{R}_{n}(y),
\end{eqnarray}
where
$\hat{\rho}_{n}(y_{0}) = \hat{\psi}_{n}^{\dag}(y_{0}) \hat{\psi}_{n}(y_{0})$ 
is the electron density in the $(n, y_{0})$ space, 
$W'(y_{0}) \equiv \partial_{y_{0}}W(y_{0})= eE_{y}(y_{0})$ and 
\begin{eqnarray}
{\cal M}_{n}(y) &=& - \{1/(2e_{n})\}\,  {\cal D}_{N}(y), \ \ 
{\cal M}_{n=0_{\pm}}(y) =0, 
\nonumber\\
 {\cal D}_{N}(y) &=&  \sum_{m=0}^{N-1} |\phi_{m}(y)|^2, \ {\cal D}_{0}(y) = 0,
\nonumber\\
\hat{R}_{n}(y) &=& R_{n}(y) + (\ell \partial_{y})^2\, {\cal D}_{N}(y),
\nonumber\\ 
R_{n}(y) 
&=& c^{n} c^{n} \,|\phi_{N}(y)|^2 +  b^{n} b^{n} \, |\phi_{N-1}(y)|^2,
\nonumber\\
&=& \textstyle{{1\over{2}}} \{|\phi_{N}(y)|^2 +  |\phi_{N-1}(y)|^2 \} +O(\mu/e_{n});\ \  
\end{eqnarray}
${\cal D}_{N}(y)$ follows from Eq.~(\ref{Fourier_Ln}) with 
$L_{N}(x) \rightarrow L_{N-1}^{1}(x) = \sum_{n=0}^{N-1} L_{n}(x)$.
${\cal D}_{N}(y)$ and $R_{n}(y)$ are localized at $y=0$ with spread of $O(\ell)$.

The real-space distribution of current $j_{x}(y)$ is obtained by taking an expectation value 
$\langle j_{x}(y) \rangle$ for filled levels.
Each filled level $n$ is characterized by a filled domain 
$\{y_{0};  y_{0;n}^{-} \le y_{0} \le y_{0;n}^{+}\}$  in the center space $\{y_{0}\}$
and the boundary positions $y_{0;n}^{\pm}$ are fixed from the spectrum 
$\epsilon_{n}[y_{0;n}^{\pm}]  \approx \epsilon_{n}+ W(y_{0;n}^{\pm}) = \epsilon_{\rm F}$
for a given value of the Fermi energy $\epsilon_{\rm F}$.
Each filled domain has a constant density 
$\langle \hat{\rho}_{n}(y_{0}) \rangle/L_{x} = \bar{\rho} =1/(2\pi \ell^2)$.

A portion of  the current $j_{x}(y)$, associated with magnetization $\propto \hat{m}_{n} \sim {\cal M}_{n}(y)$,
is a microscopic $\lq\lq$circulating" current $j^{\rm (c)}(y)$.
Indeed, it survives, upon integration over $y_{0}$, only along the periphery of domain [I],
$y \sim y_{0;n}^{\pm}$ of each filled level $n \ge 1$,
with 
$\langle j^{\rm (c)}(y) \rangle =\sum_{n} \langle j_{n}^{\rm (c)}(y) \rangle$
and 
\begin{equation}
\langle j_{n}^{\rm (c)}(y) \rangle 
= {e\omega_{c}\over{2\pi}} {1\over{2 e_{n}}} 
\Big\{  {\cal D}_{N}(y-y_{0;n}^{+}) -  {\cal D}_{N}(y-y_{0;n}^{-})\Big\},
\label{jc_local} 
\end{equation}
sharply peaked around $y \sim y_{0;n}^{\pm}$.
Each $j_{n}^{\rm (c)}(y)$ carries a total amount
\begin{equation} 
J_{n}^{{\rm (c)};+} =   \int dy\, \langle j_{n}^{\rm (c)}(y) \rangle 
=  {e\omega_{c}\over{2 \pi}} \,{N\over{2e_{n}}}= -\bar{\rho}\,\hat{m}_{n} >0
\end{equation}
(at $y \sim y_{0;n}^{+}$) and leaves orbital magnetization 
$M_{n}^{z} = \bar{\rho}\,\hat{m}_{n} <0$ inside domain [I]. 
The current profiles $\langle j_{n}^{\rm (c)}(y) \rangle$ at $y \sim y_{0;n}^{\pm}$
are essentially fixed by orbitals $n$ and field $B$. 
They are thus insensitive to $\epsilon_{\rm F}$ and 
common to valleys $(K, K')$ for $n\ge 1$ while $j^{\rm (c)}(y)=0$ for $n=0_{\mp}$.

Another portion of $j_{x}(y)$ is essentially a Hall current, 
or a $\lq\lq$drift" current $j^{\rm (d)}$, driven by the edge potential $W'(y_{0}) = e E_{y}(y_{0})$, 
with distribution 
\begin{eqnarray}
\langle j_{n}^{\rm (d)}(y) \rangle 
&=& -{e\over{2 \pi}} \int^{y_{0;n}^{+}}_{y_{0;n}^{-}}\! dy_{0}\, W'(y_{0})\, \hat{R}_{n}(y-y_{0})
\label{jd_local}
\end{eqnarray}
for each filled level $n$. 
The density $\hat{R}_{n}(y-y_{0})$ is localized about $y_{0} \sim y$ 
and, when integrated in $y_{0}$ over a few magnetic lengths about $y$, 
amounts to unity.
Accordingly, away from the edges $y \approx y_{0;n}^{\pm }$, 
\begin{equation}
\langle j_{n}^{\rm (d)}(y)  \rangle 
= -(e/2 \pi)  W'(y) + O(W''') 
\label{jby_Wlocal}
\end{equation}
simply follows a local edge field $W'(y) = e E_{y}(y)$ 
towards the two boundaries $y \approx y_{0;n}^{\pm}$ and rapidly vanish there. 
Its amount 
$J^{\rm (d)}_{n} = \int dy\, \langle j_{n}^{\rm (d)}(y)\rangle$ per edge, 
\begin{equation} 
J_{n}^{{\rm (d)};\pm} = \mp {e\over{2 \pi}}\,  W(y_{0;n}^{\pm }) 
\approx   \mp {e\over{2 \pi}}\, (\epsilon_{\rm F} -\epsilon_{n}),
\label{jnb_GR}
\end{equation}
is essentially fixed by an edge potential $\epsilon_{\rm F} - \epsilon_{n} >0$
acting on each level $n$ in each edge region $\sim y_{0;n}^{\pm}$.

\section{P-N junction in graphene}

In this section we consider edge-current distributions along  a $p$-$n$ junction.
Let us again look at the Hall-bar sample in Fig.~1 and 
suppose that the wall $W(y) = -e A_{0}(y)$ gently rises around $y\sim 0$ 
and reaches a height $W_{\rm h}$ deep in domain [II], $y\gg 0$. 
The wall around $y\sim 0$ forms a junction of active electrons and holes in adjacent domains 
when $\epsilon_{\rm F}$ is set to cross the wall.

 \begin{figure}[tpb]
\begin{center}
\includegraphics[scale=0.84]{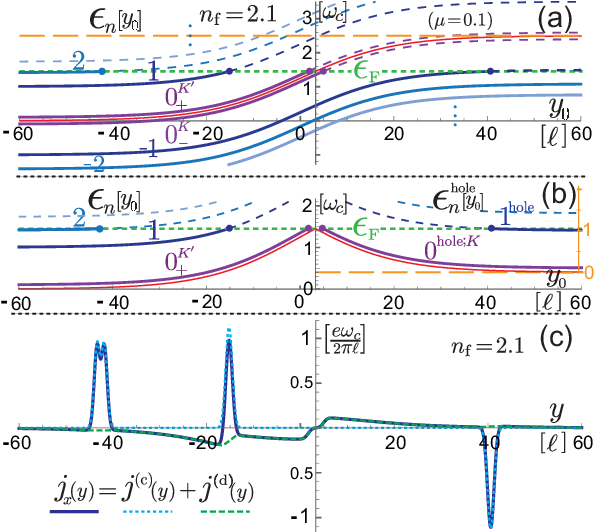}
\end{center}
\vskip-.7cm
\caption{
P-N junction in graphene.
(a)~Level spectra of electrons across the $p$-$n$ junction at $n_{\rm f} =2.1$, with
wall height $W_{\rm h}= 2.5\,  \omega_{c}$ and valley breaking $\mu=0.1$.
A thin red curve indicates the charge-neutral surface of the Dirac sea. 
(b)~Electron-hole picture of the junction. 
(c)~Current distributions along the junction in adjacent electron and hole domains.
In each domain two species of persistent current 
$j^{\rm (c)}(y)$ and $j^{\rm (d)}(y)$ mutually flow in opposite directions 
and their distributions  are in one-to-one correspondence with the electron and hole spectra. 
}
\end{figure}

 \begin{figure}[ptb]
\begin{center}
\includegraphics[scale=0.48]{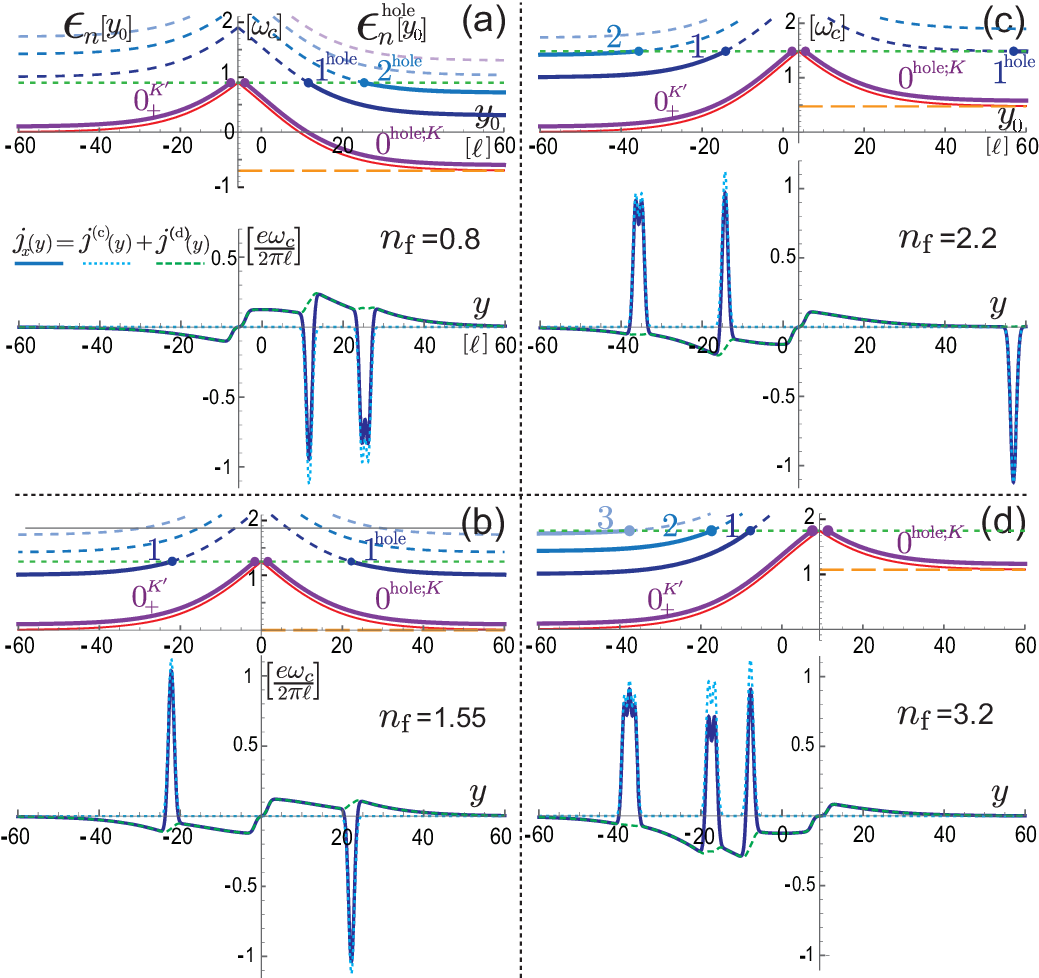}
\end{center}
\vskip-.7cm
\caption{ 
(a) - (d)~Electron and hole spectra across the junction 
and associated edge current distributions 
show a characteristic change as $\epsilon_{\rm F} \sim n_{\rm f}$ is varied 
within the wall height $W_{\rm h}$. 
}
\end{figure}

For numerical simulations we adopt a potential of the form 
\begin{equation}
W(y) = W_{\rm h}\, \{1 + {\rm tanh}(\lambda\,  y/\ell) \}/2,
\label{PotWall}
\end{equation}
with  height $W_{\rm h}= 2.5\,  \omega_{c}$ and $\lambda=1/20$.
As seen from Fig.~2(a),
it accommodates the edge modes 
of level $n=(0_{+}^{K'}, 1^{K,K'}, 2^{K,K'},\cdots)$ 
over the range $- 50\ell  \lesssim y_{0}  \lesssim 0$ in domain [I] 
while it supports empty electron modes (or "hole" modes) 
of $n=(0_{-}^{K}, -1^{K,K'}, -2^{K,K'},\cdots)$ in adjacent domain [II].

We write $\epsilon_{\rm F} =\omega_{c}\sqrt{n_{\rm f}+\mu^2}$ 
and use $n_{\rm f}$ to specify filling of the edge modes.
We now take the electron spins $(\downarrow\uparrow)$ into account 
while ignoring their splitting. 
Accordingly,   $0< n_{\rm f} <1$ refers to filling of 
the $n=0_{+}^{K'}|_{\downarrow \uparrow}$ levels near the total filling factor $\nu = 2$, 
$1< n_{\rm f} <2$ to filling of four $n=1^{K, K'}|_{\downarrow \uparrow}$ levels near $\nu = 6$, 
$2< n_{\rm f} <3$ to filling of four $n=2$ levels near $\nu = 10$, etc.

Figure~2(a) shows the level spectra across the junction at $n_{\rm f} =2.1$, 
with somewhat larger valley breaking $\mu =0.1$ taken  
for better visibility of the $n=(0^{K}_{-}, 0^{K'}_{+})$ levels.
A practical way to visualize currents along the $p$-$n$ junction is 
to pass to an electron-hole picture of the junction. 
Folding back the $(y_{0} \gtrsim 0)$ spectra vertically 
at the Fermi level $\epsilon_{\rm F}$ 
leads to the spectra of active electron and hole levels in separate domains, 
depicted in Fig.~2(b).  
The empty portion of electron level $n=-N$ of energy 
$\epsilon_{-N}[y_{0}] =\epsilon_{-N} + W(y_{0})$ has thereby turned 
into holes of level $n^{\rm hole} =N$ of energy 
\begin{equation}
{\cal E}_{N}^{\rm hole}[y_{0}] = \epsilon_{N}  - W(y_{0}) +W_{\rm h}
= \epsilon_{N}[y_{0}]|_{-A_{0}}+ W_{\rm h},\ \ 
\end{equation}
now measured relative to the interior $(y_{0} \gg 0)$ of the hole domain  
at energy ${\cal E} = 2\epsilon_{F} -W_{\rm h}$, where no current survives;
${\cal E}^{\rm hole}_{N}[y_{0}] \rightarrow \epsilon_{N}$  for $y_{0} \gg 0$.
The inverted wall $-W(y_{0})$ now serves to confine holes of level 
$0_{+}^{{\rm hole};K}$ (of spectrum ${\cal E}_{0_{+}}^{K'}|_{-A_{0}} + W_{\rm h}$) 
and of level $1^{{\rm hole}; K, K'}$. 
This is in accord with Eq.~(\ref{hole_e_spectra}).

Currents carried by active electrons and holes are readily read from the level spectra 
since there is one-to-one correspondence between them, 
as noted in Eqs.~(\ref{jc_local}) - (\ref{jby_Wlocal}).
Figure~2(c) presents the edge-current distribution
$\langle j^{\rm (c)}(y) \rangle + \langle j^{\rm (d)}(y) \rangle$
along the junction at $n_{\rm f}=  2.1$. 
The circulating currents $\langle j_{n}^{\rm (c)}(y) \rangle$ are sharply peaked 
about the edge boundaries $y \sim y_{0;n}^{\pm}$ for $n \ge 1$
while $j_{n _{}= 0_{\pm }}^{\rm (c)}  =0$. 
In contrast, the drift currents $\langle j_{n}^{\rm (d)}(y) \rangle$ 
gradually grow with a broad profile $\propto \pm W'(y)$ towards the junction 
and vanish rapidly around $y \sim y_{0;n}^{\pm}$.

In equilibrium  two species of persistent current $j^{\rm (c)}$ and $j^{\rm (d)}$ 
flow in opposite directions along the periphery of a populated domain. 
For both electrons and holes, $j^{\rm (c)}$ is diamagnetic in nature 
 while $j^{\rm (d)}$ is paramagnetic.
 This is why, in Fig.~2(c),
$\langle j_{n}^{\rm (c)}(y) \rangle$ and $\langle j_{n}^{\rm (d)}(y) \rangle$ 
both change sign across the junction.
The electrons and holes circulate or drift in the same direction along the junction on each side.

Figure 3 illustrates how the $e$-$h$ spectra and associated current distributions change
across the junction with filling $\sim n_{\rm f}$.
For $0 < n_{\rm f} \lesssim 1$ only  the drift current $j^{\rm (d)}_{0_{+}}$ 
of $n=0_{+}^{K'}$ arises on the electron side 
while a variety of current channels is active on the hole side. 
With increasing $n_{\rm f}$,  active level spectra change and,
on the electron side,  circulating components $j_{n}^{\rm (c)}$ appear one after another 
at intervals $\Delta n_{\rm f} \approx 1$ 
over a broad background of $j^{\rm (d)}$ of opposite polarity,
thus forming an alternating pattern of current channels.
On increasing $n_{\rm f}$ these current channels shift toward the junction 
while, on the hole side, a similar pattern of current channels goes away, 
with $(j_{n}^{\rm (c)}, j_{n}^{\rm (d)})$ disappearing one by one.
Such behavior of persistent currents is in qualitative agreement with observations~\cite{UKBL}.

\section{Coulomb interaction} 

In this section we study many-body effects on current distributions.
The Coulomb interaction is denoted as
\begin{equation}
V_{c}[\rho] = {1\over{2}} \sum_{\bf p} v^{\rm C}_{\bf p} \rho_{\bf -p}\, \rho_{\bf p} ,
\end{equation}
with the potential $v^{\rm C}_{\bf p}= 2\pi \alpha_{e}/(\epsilon_{\rm b} |{\bf p}|)$,
$\alpha_{e} \equiv e^{2}/(4 \pi \epsilon_{0})$ and 
the substrate dielectric constant $\epsilon_{\rm b}$.
The electron density 
$\rho_{-{\bf p}} =\int d^{2}{\bf x}\,  e^{i {\bf p\cdot x}}\, \Psi^{\dag} \Psi$ 
is rewritten as
\begin{equation}
\rho_{-{\bf p}} =  [e^{i{\bf p\cdot X}}]^{mn} 
\int\! dy_{0}\, \psi^{\dag}_{m}(y_{0})\,e^{i{\bf p}\cdot {\bf r}}\psi_{n}(y_{0}),
\label{U_p}
\end{equation}
where $[e^{i{\bf p\cdot X}}]^{mn} = \gamma_{\bf p}\, g^{mn}_{\bf p}$.

In terms of the one-body eigenmodes $\hat{\psi} \equiv U\psi$, 
$\rho_{-{\bf p}} = \hat{\rho}_{-{\bf p}} + \delta \hat{\rho}_{-{\bf p}}$
and $V_{c}[\rho] =V_{c}[\hat{\rho}]+ \delta V_{c}$ 
acquire modifications of $O(v)$,  $O(vA'_{0})$, etc.; 
$\hat{\rho}_{\bf p}$ stands for $\rho_{\bf p}$ with $\psi \rightarrow \hat{\psi}$.
Actually, $\delta \hat{\rho}_{-{\bf p}}$ has an $O(v'_{x})$ modification of the form
\begin{equation}
\delta^{b} \hat{\rho}_{\bf -p} =
 - i\,  {1\over{2}} \int \! dy_{0}\, b_{z}(y_{0})\, 
 \hat{\psi}^{\dag}(y_{0})\,({\bf p} \cdot\! {\bf X})\,  e^{i{\bf p\cdot x}} \hat{\psi}(y_{0}).
\end{equation}
For active electrons (or holes) of level $n$ in the bulk (i.e., away from the $p$-$n$ junction),
this  $\delta^{b} \hat{\rho}_{\bf -p}$, via exchange interaction in $\delta V_{c}$,  
works to slightly reduce~\cite{ks_edgeC_W}
the orbital moment, 
$\hat{m}_{n} = - e \ell^2 \omega_{c} N/(2 e_{n}) \rightarrow \hat{m}_{n}  + \Delta \hat{m}_{n}$,
with 
\begin{equation}
\Delta \hat{m}_{n} = e\ell^2 \sum_{m} Q^{mn}_{0}, 
\label{DMn}
\end{equation}
where the sum is taken over filled levels $m\in \{0_{+}, 1, 2, \cdots\}$ and 
$Q^{mn}_{0} = Q^{nm}_{0}$ are given by 
\begin{equation}
(Q^{00}_{0}, Q^{10}_{0}, Q^{11}_{0})
= \tilde{V}_{c}\,({1\over{4}}, {1\over{16}}, {11\over{64}}), \ \ 
 \tilde{V}_{c }= {\alpha_{e}\over{\epsilon_{b}\, \ell}} \sqrt{\pi\over{2}},
\end{equation}
etc., for $\mu \rightarrow 0$.
It thus also acts to somewhat reduce the amount of circulating currents $J_{n}^{\rm (c)}$.
Note that $\Delta \hat{m}_{0} >0$.  This implies that the $n=0_{+}^{K'}$ level, 
when filled, now supports a weak circulating component
$j_{0_{+}}^{\rm (c)}$ of many-body origin, 
which flows in the same direction as the drift one $j_{0_{+}}^{\rm (d)}$ 
and thus can easily escape detection.

On the other hand, $\delta V_{c}$ acquires two types of  $O(v_{x}A'_{0})$ terms, 
which essentially combine to vanish at long wavelengths~\cite{ks_edgeC_W}. 
This means that the Coulomb interaction does not affect the total drift currents 
$J_{n}^{\rm (d)}$, which are thus governed by 
the one-body formula in Eqs.~(\ref{jnb_GR}).
One still has to note that the level spectra and associated edge boundaries 
$y_{0;n}^{\pm}$ generally have corrections of $O(\tilde{V}_{c})$.

 \begin{figure*}[tbp]
\begin{center}
\includegraphics[scale=1.75]{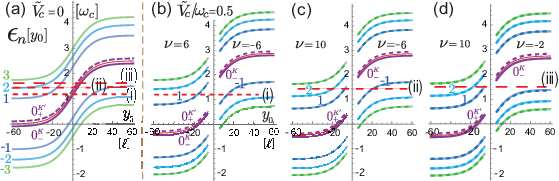}
\end{center}
\vskip-.7cm
\caption{
Level spectra $\epsilon_{n}[y_{0}]$ across the $p$-$n$ junction 
in (a) the absence and (b)-(d) presence of the Coulomb interaction, 
with $\tilde{V}_{c}/\omega_{c} = 0.5$ and $\mu = 0.1$ chosen.
Solid curves refer to valley $K$ and dashed ones to $K'$; 
valley splitting is almost invisible, except for $0_{-}^{K}$ and $0_{+}^{K'}$.
$\nu$ stands for the filling factor in the bulk $y_{0} \ll 0$ or $y_{0} \gg 0$, 
with the spin degree of freedom included.
Dashed lines (i)-(iii) refer to some choices of $\epsilon_{\rm F}$, 
contrasting the $\tilde{V}_{c}=0$ and $\tilde{V}_{c}\not=0$ cases.
}
\end{figure*}

For electrons of level $n$ in the bulk, 
$V_{c}[\hat{\rho}]$ leads, via exchange interaction, 
to self-energies of the form 
\begin{eqnarray}
\Delta \epsilon_{n} = - \sum_{\bf p} v_{\bf p} \gamma_{\bf p}^{2}  
\sum_{m: {\rm filled}} |g_{\bf p}^{nm}|^{2},
\end{eqnarray}
where $\sum_{m}$ includes also a sum over filled levels $m$ in the Dirac sea.
Actually one can write, noting $\gamma_{\bf p}^{2}\sum_{m}|g^{n m}_{\bf p}|^2 =1$, 
a half-infinite sum in the form $\gamma_{\bf p}^{2}\sum_{m \le -1}|g^{n m}_{\bf p}|^2 
= -{1\over{2}} +  {1\over{2}}{\cal F}_{n}({\bf p}^2;\mu) + \cdots$, 
where 
\begin{equation}
{\cal F}_{n}({\bf p}^2;\mu) \equiv
\gamma_{\bf p}^{2}\sum_{k=1}^{\infty}\{ |g^{nk}_{\bf p}|^2 -  |g^{n,-k}_{\bf p}|^2 \},
\end{equation}
and safely eliminate the $-{1\over{2}}$ term, common to all levels $n$,
by adjusting the zero of energy. 
Infinite sums ${\cal F}_{n}({\bf p}^2;\mu)$ represent 
many-body corrections coming from the Dirac sea.
They actually are afflicted with ultraviolet divergences, 
which are handled by renormalization of the velocity $v$ and valley gap $\mu$. 
They thereby are replaced by renormalized expressions 
${\cal F}^{\rm ren}_{n}({\bf p}^2;\mu)$ and 
the renormalized self-energies are written as
\begin{equation}
\Delta \epsilon_{n}^{\rm ren} = \Omega_{n}(\mu) 
- \sum_{k} \nu [k]\,  \sum_{\bf p}v_{\bf p} \gamma_{\bf p}^{2}  |g^{nk}_{\bf p}|^2,
\label{En_reg}
\end{equation}
with $\Omega_{n}(\mu) \equiv \sum_{\bf p}v_{\bf p}\,
{\textstyle{1\over{2}}}{\cal F}^{\rm ren}_{n}({\bf p}^2;\mu)$. 
The filling factor, 
\begin{equation}
\nu[k] = \nu_{k}\, \theta_{(k \ge1)} -  (1-\nu_{k})\theta_{(k\le-1)}
 + (\nu_{0}- {\textstyle{1\over{2}} })\, \delta_{k,0},
 \end{equation}
 with $\theta_{(k\ge1)} =1$ for $k\ge1$ and $\theta_{(k\ge1)} =0$ 
 otherwise, etc.,
refers to a finite number of 
electron levels $k$ (with filling $0 \le \nu_{k} \le 1$) 
around the $\nu=0$ vacuum state.
See Ref.~\cite{ks_CR_GR} for details of the renormalization procedure.
In this form self-energies $\Delta \epsilon_{n}^{\rm ren}$ 
enjoy the $e$-$h$ conjugation property in Eq.~(\ref{EKandEKp}).   
Appendix B summarizes some explicit forms
of $\Omega_{n}(\mu)$ to be used in numerical simulations below.

For simulations let us take $\tilde{V}_{c}/\omega_{c} = 0.5$ and $\mu=0.1$.
It is in general difficult to determine the level spectra across the junction, 
since $\Delta \epsilon_{n}^{\rm ren}$ sensitively depend on the filling  of levels around it. 
In contrast, away from the junction $(y_{0}\ll 0$ and $y_{0}\gg 0)$ 
the spectra take a simple form 
$\epsilon_{n} + W(y_{0}) + \Delta \epsilon_{n}^{\rm ren}$.

Figure~4(b) shows such spectra, those
with levels $(0_{+}^{K'}, 1^{K,K'})$ filled for $y_{0 }\ll 0$  (at filling $\nu \sim 6$)
and $(0_{-}^{K}, -1^{K,K'})$ left empty  (at $\nu\sim -6$) for $y_{0} \gg 0$
so that $\epsilon_{\rm F}$ lies somewhere around dashed line (i).
Interpolating them smoothly across the junction allows one to speculate the real spectra there.
One can then identify, for a given filling $\sim \epsilon_{\rm F}$, 
possible configurations of edge boundaries $\{ y_{0;n}^{\pm}\}$ 
and associated currents $\{ j_{n}^{\rm (c)}(y),  j_{n}^{\rm (d)}(y) \}$. 
[For comparison, Fig.~4(a) shows the $\tilde{V}_{c}=0$ spectra.]
In particular, at $\epsilon_{\rm F}$ indicated by dashed line (i), 
current channels of electron levels $(0_{+}^{K'}, 1^{K,K'})$ 
and hole levels $ (0^{{\rm hole}; K}, 1^{{\rm hole}; K,K'})$ arise
in a configuration similar to those in Fig.~3(b) 
of the $\tilde{V}_{c}=0$ case at $n_{\rm f} = 1.55$.

As $\epsilon_{\rm F}$ is raised to the dashed line (ii) in Fig.~4(c), 
the electron levels $2^{K,K'}$ come into play and the Coulomb-corrected level spectra 
change accordingly  ($\nu \rightarrow 10$) on the electron side, 
realizing a current configuration similar to one in Fig.~3(c) at $n_{\rm f} \sim 2.2$.
At a slightly higher filling [dashed line (iii)] in Fig.~4(d), 
a prominent profile of $j^{\rm (c); hole}_{1}(y)$ disappears, 
leaving only the drift current $j^{\rm (d); hole}_{0}$ on the $(\nu \rightarrow -2)$ hole side.

It is clear from Fig.~4 that the Coulomb interaction 
effectively enhances the junction height $W_{\rm h}$. 
Valley splitting [in $(K,K')$] of level spectra is almost invisible, 
except for the $N=0$ sector $\{0_{-}^{K}, 0_{+}^{K'}\}$.  
When Coulomb-enhanced spin or valley gaps develop around some integer fillings 
on the electron and/or hole sides, 
the $p$-$n$ junction will display a variety of edge-current configurations. 
In this sense, $p$-$n$ junctions, via detection of associated microscopic currents,
will be a useful ground for investigating many-body quantum effects in graphene.

 \section{summary and discussion}
 
A gate-induced $p$-$n$ junction in graphene manifests 
the particle-hole character of Dirac electrons. 
The junction contrasts the edge states of electrons and holes on each side of it. 
In a magnetic field, those edge states support two species of current, 
a diamagnetic circulating current $j^{\rm (c)}$ and
a paramagnetic drift current $j^{\rm (d)}$, on each side of the junction. 
In this paper we have examined real-space distributions of those persistent currents 
along the junction, i.e., how $j^{\rm (c)}$ and $j^{\rm (d)}$ emerge in alternating pattern 
and move on or away on each side
as the Fermi level $\epsilon_{\rm F}$ is gradually varied in equilibrium.

The spectra and response of positive- and negative-energy electrons, 
or those of electrons and holes, 
are essentially governed by the $e$-$h$ conjugation property of the Dirac Hamiltonian.
The edge currents along the junction on each side are directly read 
from the electron and hole spectra over the filled Dirac sea, as seen from Fig.~2.
It is crucial for such a simple electron-hole picture of persistent currents in graphene 
that the ground state, the Dirac sea filled with negative-energy electrons, 
as a whole remains electrically inactive. 
Magnetically, in contrast, the filled sea shows intrinsic orbital diamagnetism of many-body origin.

In this way, the $p$-$n$ junction in graphene serves to project the edge-state spectra 
onto the real space in the form of current distributions, 
with resolutions controllable by the gradient and height of the junction.
Now that local currents are observable in experiment by use of a nanoscale magnetometer,
$p$-$n$ junctions in graphene, few-layer graphene and related atomic layers 
will be a useful platform 
for exploring intriguing many-body and topological quantum phenomena.

\acknowledgments
 This work was supported in part by JSPS KAKENHI Grant No. 21K03534.
 
\appendix

\section{diagonalization}
In this appendix  we outline diagonalization of the one-body Hamiltonian ${\cal H}^{mn}$ 
in Eq.~(\ref{H_gr}) to $O(v)$,  $O(A_{0})$ and $O(vA_{0})$. 
Let us adopt static probes $(v, A_{0})$ and 
set  ${\cal H}^{mn} = \epsilon_{n} \delta^{mn} + V^{mn}$
and $V^{mn}= V_{v}^{mn} + V_{A}^{mn}$, with 
 \begin{eqnarray}
V_{v}^{mn} &=& i\omega_{c}  \{ -b^{m}[v]^{M-1,N}c^{n} + c^{m}[v^{\dag}]^{M, N-1}b^{n} \},
\nonumber\\
V_{A}^{mn}/e &\approx&  - \gamma_{\bf p}A_{0} \delta^{mn} +
 (Z^{\dag})^{mn} \ell  E + Z^{mn}  \ell  E^{\dag}, 
\end{eqnarray}
where $E = (E_{y} + iE_{x})/\sqrt{2} \rightarrow - \partial_{y_{0}} A_{0}(y_{0})/\sqrt{2}$;
$Z^{mn} = c^{n}Z^{MN}c^{n} + b^{m}Z^{M-1,N-1}b^{n}$.
Here we retain only terms up to $O(\partial A_{0})$ 
for $A_{0}$, assuming its gentle  variations.

Upon diagonalization to $O(V^2)$, 
${\cal H}^{mn}$ turns into 
\begin{equation}
\hat{\cal H}^{nn} = \epsilon_{n}  + V^{nn} + \sum_{k} V^{nk}V^{kn}/\epsilon_{nk},
\end{equation}
where $\epsilon_{nk} = \epsilon_{n}- \epsilon_{k}$.
The response of $O(v)$ and $O(A_{0})$ follows from $V^{nn}$.

The $O(v\partial A_{0})$ response is written as 
\begin{equation}
\hat{V}^{(vA)}_{n}
=  e\ell ( P^{\dag}_{n} E +E^{\dag} P_{n}),
\end{equation}
where $P_{n} = P_{n}^{(-)} + P_{n}^{(+)}$ and
 \begin{equation}
 P_{n}^{(-)} = \sum_{k} {1\over{\epsilon_{nk}}} V_{v}^{nk}Z^{kn},
P_{n}^{(+)} =\sum_{k} {1\over{\epsilon_{nk}}}  Z^{nk}V_{v}^{kn}.
\end{equation}
The sum over intermediate levels $\{k\}$ is neatly carried out 
with the aid of the formulas
 \begin{eqnarray}
&&\sum_{k=\pm K}(e_{n}+e_{k})c^{k}b^{k} = -\sqrt{K},
\\
&&\sum_{k=\pm K}(e_{n}+e_{k})\{c^{k}c^{k} , b^{k}b^{k} \} = \{e_{n}- \mu, e_{n}+\mu\},\ \ 
\\
&&(e_{n}-\mu) b^{n} = -\sqrt{N} c^{n},   
(e_{n} + \mu) c^{n} = -\sqrt{N} b^{n},
\end{eqnarray}
where $n = \pm N$. 
Actually, $P^{(-)}_{n}$ consists of a sum over $k= \pm K$ with $K= N-1$,
and $P^{(+)}_{n}$ with $K= N+1$.
The result is  
 \begin{eqnarray}
P_{n}^{(-)} \! \!\! 
&=&\! i \{ (N\! -b^{n}b^{n})  [v]^{N-1,N-1} \!- \! \sqrt{N(N\!-\!1)} [v^{\dag}]^{N,N-2} \} 
\nonumber\\
&=&\! i \{ (N\!-b^{n}b^{n}) L_{N-1}(\xi) u - L_{N-2}^{2}(\xi) (\ell\partial)^{2}u^{\dag}  \}, 
\\
P_{n}^{(+)}\!\!\! 
&=&\! i \{- (N\!+c^{n}c^{n})  [v]^{NN}\! +\! \sqrt{N(N\!+\!1)} [v^{\dag}]^{N+1,N-1} \}
\nonumber\\
&=&\! i \{- (N\! + c^{n}c^{n}) L_{N}(\xi) u + L_{N-1}^{2}(\xi) (\ell \partial)^{2}u^{\dag}  \},
\end{eqnarray}
where $u \equiv \gamma_{\bf p} v(y_{0})$. 
This $P_{n}$ leads to the $O(v E)$ response in Eq.~(\ref{Vnr}),
 $\hat{V}^{(vA)}_{n} = e \ell v_{x}(y_{0}) E_{y}(y_{0}) + \cdots$,  
which implies that each filled level $n$ supports a Hall current
of one unit of conductance $\sigma_{xy} =- e^2\ell^2 \bar{\rho} =- e^2/(2\pi \hbar)$.

Finally, to handle the Dirac sea carefully, let us suppose 
that each valley ($K$ or $K'$) consists of $(2N_{\rm cut} +1)$ Landau levels $\{n\}$ 
with $-N_{\rm cut} \le n \le N_{\rm cut}$.
Then at the bottom level $n= -N_{\rm cut}$  there is no $P^{(+)}_{n} (\rightarrow 0)$, 
otherwise charge will be lost via $-N_{\rm cut } \leftrightarrow k=\pm (N_{\rm cut}+1)$ mixing.
Keeping only $P^{(-)}_{n}$ yields (for $N_{\rm cut} \rightarrow \infty$) 
\begin{equation}
\hat{V}^{(vA)}_{n= -N_{\rm cut}} = e\ell E_{y} \{ - (N_{\rm cut}-1/2)  v_{x} + O(\nabla^2) v_{x} \}.
\end{equation} 
The bottom level thus supports $-(N_{\rm cut}-1/2)$ units of Hall current.
[For filled levels $\{n\}$ (with uniform density $\bar{\rho}$) 
deep in the Dirac sea,
the $O(\nabla^2) v_{x}$ terms make no contribution of $O(E)$ to the current $j_{x}$.]
Accordingly, 
the filled sea of valley $K$ with $n= (0_{-}, -1, \dots, -N_{\rm cut})$, 
 as a whole, supports only $1/2$ unit of $\sigma_{xy} = -e^2/h$ 
and that of valley $K'$ shares $-1/2$ unit. 
No vacuum current eventually remains for the full Dirac sea of $K+K'$.

\section{Intrinsic magnetization of the Dirac sea}

In this appendix we summarize some expressions related to intrinsic magnetization 
of the vacuum state.
Let us first note the asymptotic formula 
\begin{equation}
\sum_{n=1}^{N}n^{z} \approx {1\over{z+1}} (N + 1/2)^{z +1} + \zeta (-z)
\end{equation}
for $z <1$ and $N \rightarrow \infty$.
Direct calculations of vacuum energy densities 
$\epsilon_{\rm vac}^{B}$ and $\epsilon^{B=0}_{\rm vac}$,  
noted in Eq.~(\ref{E_B_vac}), then yield (for $N_{\rm cut} \rightarrow \infty$)
\begin{eqnarray}
\epsilon_{\rm vac}^{B}&=&\textstyle  - \bar{\rho}\,  \omega_{c} 
\Big\{ {4\over{3}} \bar{N}^{3/2} + 2\, \zeta (- {1\over{2}})
+ |\mu| + O(\mu^2)  \Big\},
\nonumber\\
\epsilon^{B=0}_{\rm vac}&=&\textstyle - {4\over{3}} \bar{\rho}\,  \omega_{c} 
\Big\{ ({1\over{2}}\ell^{2} k_{\rm F}^2 + \mu^2)^{3/2} - |\mu|^{3} \Big\},
\end{eqnarray}
where $\bar{N} \equiv N_{\rm cut} + {1\over{2}} + \mu^2 = {1\over{2}}\ell^{2} k_{\rm F}^2 + \mu^2$. 
This leads to the vacuum energy 
$\epsilon_{\rm vac} =\epsilon^{B}_{\rm vac} -\epsilon^{B=0}_{\rm vac}$ 
quoted in Eq.~(\ref{E_vac}).

\section{Self-energy corrections}

Equation~(\ref{En_reg}) refers to renormalized self-energy corrections cast in the form 
$\Delta \epsilon_{n}^{\rm ren} = \Omega_{n}(\mu)  - \sum_{m}\nu[m] \, T^{nm}$,
with
$T^{mn} \equiv \sum_{\bf p} v_{\bf p}  \gamma_{\bf p} ^{2} |g^{nm}_{\bf p} |^{2} = T^{nm}$.
In this appendix we summarize a set of $\{ \Omega_{n}, T^{nm}\}$, 
used to draw Fig. 4.

The  many-body corrections $\{\Omega_{n}\}$ have the $e$-$h$ conjugation property 
$\Omega_{-n} = -\Omega_{n}|_{-\mu}$ and 
$\Omega_{n}^{K'} = \Omega^{K}_{n}|_{-\mu}$. 
Some of them, quoted from Ref.~\cite{ks_CR_GR}, 
read
\begin{eqnarray}
\Omega_{0_{-}}^{K}\!\! &=& -\Omega_{0_{+}}^{K'} \approx \tilde{V}_{c}\,  (0.01688 \mu), 
\nonumber\\
\Omega_{1}^{K} &\approx& \tilde{V}_{c}\, ( {\textstyle{3\over{8}}} + 0.0887 \mu),\ 
\Omega_{2}^{K} \approx \tilde{V}_{c}\, ( 0.4074 + 0.0520 \mu),
\nonumber\\
\Omega_{3}^{K} &\approx& \tilde{V}_{c}\, ( 0.4054 + 0.0370 \mu).
\end{eqnarray}

Similarly, $\{ T^{mn} \}$ enjoy the feature 
$T^{mn}|^{K'} = T^{mn}|^{K}_{-\mu}= T^{-m,-n}|^{K}$.
Some basic ones read 
\begin{eqnarray}
T^{0_{-}0_{-}} \!\! &=& \tilde{V}_{c}, \ 
T^{\pm 1,0_{-}} \approx \tilde{V}_{c}\,  {\textstyle{1\over{4}}}\,  (1\mp  \mu),
\nonumber\\ 
T^{\pm 2,0_{-}}\!\! &\approx& 
\tilde{V}_{c}\,  \textstyle {3\over{16}}\, (1\mp {1\over{\sqrt{2}}}\, \mu),
\nonumber\\
T^{11}   &\approx& \tilde{V}_{c}\,  \textstyle {1\over{16}} (11 + 2\mu), \ 
T^{-1,1}   \approx  \tilde{V}_{c}\, {3\over{16}},
\nonumber\\
T^{\pm 2,1}  &\approx& 
\tilde{V}_{c}\,  \textstyle{1\over{64}} \{15+ \mu  \pm (4\sqrt{2} + {1\over{\sqrt{2}}} \mu) \}.
\end{eqnarray}




\begin{thebibliography}{99}
\bibitem{fn} 
 In the present paper we have confirmed the result of earlier calculations 
 in Ref.~\cite{ks_edgeC_W} by use of the standard diagonalization procedure alone; 
 the last term $\propto h^{nn}(\xi)$ in Eq.~(\ref{Vnr}) was thereby fixed anew.
\end{thebibliography}
\end{document}